\documentclass[final]{IEEEtran}

\usepackage{amsthm,amssymb,graphicx,multirow,amsmath,color,amsfonts}
\usepackage[update,prepend]{epstopdf}
\usepackage[noadjust]{cite}
\usepackage[latin1]{inputenc}
\usepackage{tikz}
\usetikzlibrary{arrows,calc}		
\usepackage{bbm} 
\usepackage{pdfpages}
\usepackage{tabulary}
\usepackage{multirow}
\usepackage{comment}

\usepackage{etoolbox}

\BeforeBeginEnvironment{figure}{\vskip-2ex}
\AfterEndEnvironment{figure}{\vskip-1ex}
\def\chr#1{{\color{black}#1}}
\def\chb#1{{\color{black} #1}}

 \def \zz {\theta}

\def\nb0{{\mathbf{0}}}
\def\nb1{{\mathbf{1}}}

\def\nbbR{{\mathbb{R}}}

\newtheorem{thm}{Theorem}
\newtheorem{definition}{Definition}

\newtheorem{cor}{Corollary}

\newtheorem{remark}{Remark}

\def\argmin{\operatorname{arg~min}}

\def\figref#1{Fig.\,\ref{#1}}%
\def\E{\mathbb{E}}

\def\P{\mathbb{P}}

\def\R{\mathbb{R}}

\allowdisplaybreaks 

\begin{document}
\graphicspath{{./Figures/}}
\title{ 
Nearest-Neighbor and  Contact Distance Distributions for Thomas Cluster Process
}
\author{Mehrnaz Afshang, Chiranjib Saha, and Harpreet S. Dhillon
\thanks{The authors are with Wireless@VT, Department of ECE, Virgina Tech, Blacksburg, VA, USA. Email: \{mehrnaz, csaha,  hdhillon\}@vt.edu. The support of the US NSF (Grants CCF-1464293, CNS-1617896, IIS-1633363) is gratefully acknowledged. Last updated: \today.} 
}

\maketitle
 \vspace{-1 cm}
\begin{abstract}
We  characterize the statistics of  nearest-neighbor and contact distance distributions for Thomas cluster process (TCP), which is a special case of Poisson cluster process. 
\chb{In particular, we derive the cumulative distribution function (CDF) of the distance to the nearest point of TCP from a reference point for three different cases: (i) reference point is not a part of the point process, (ii) it is chosen uniformly at random from the TCP, and (iii) it is a randomly chosen point from a cluster chosen uniformly at random from the TCP. While the first corresponds to the contact distance distribution, the other two provide two different viewpoints for the nearest-neighbor distance distribution. Closed-form bounds are also provided for the first two cases.}
\end{abstract}
\begin{IEEEkeywords}
Stochastic geometry, Thomas cluster process, contact distance, and nearest-neighbor distance.
\end{IEEEkeywords}

\section{Introduction}
Poisson cluster processes have  numerous applications  in diverse branches of science (\chb{such as geodesy and ecology}) since it captures the attraction (clustering) between nearby points, which is a common occurrence in many point patterns~\cite{chiu2013stochastic}.
In wireless \chb{networks},  the \chb{locations} of users (network subscribers) tend to be clustered. For instance,  users are typically concentrated in  some specific areas, e.g., residential and/or commercial complexes, called \chb{\em user hotspots}.  The clustering nature of user distribution is often considered in the modeling of \chb{real-world} wireless networks~\cite{SahaAfshDh2016}. For example,  3GPP customarily models the users forming  clusters over a circular disc in their simulations~\cite{3gpp2}. Moreover, subscriber-owned small cell base stations (SBSs) are typically deployed on the scale of one per household/business to serve a group (cluster) of users. Thus, the spatial distribution of the SBSs also exhibit clustering~\cite{spatialmodelingAndrews2010}.  This makes Poisson cluster process a natural choice for modeling user as well as SBS locations in wireless networks. \chb{As evident from the prior art on the analysis of wireless networks using tools from stochastic geometry~\cite{haenggi2012stochastic}, nearest-neighbor and contact distance distributions play a crucial role in the tractable characterization of key performance metrics.} \chr{However, the derivation of explicit expressions for these distributions for general Poisson cluster processes are not well explored.}
  Thus,  despite Poisson cluster process being a suitable model for SBS and user distributions, its applicability  in the analysis of wireless networks is significantly restricted.

In this letter, we focus on characterizing the   nearest-neighbor and contact distance distributions of TCP, which  is a popular  \chb{special case of}  Poisson cluster \chb{process}.  \chb{The sparsely existing prior art of  relevance  can be classified in two main  directions}.
The first one focuses on formulating likelihood function of a multinomial Poisson cluster process which yields the desired distributions for a TCP as a special case~\cite{baudin1981likelihood}. \chr{Though the results derived are general and applicable for a wide family of Poisson cluster processes, the exposition of the analysis requires substantial 	proficiency in measure theory and \chr{is hence somewhat} pedantic  to attract general interest in the wireless community. } 
%
%
\chr{ As a result, the} second direction,  which is a common practice in the literature of wireless networks, is to approximate the first order statistic of the Poisson cluster process with that of homogenous Poisson point process (PPP), and then simply use the well-known  contact distribution of PPP; see~\cite{HetPCPGhrayeb2015,Wang2016} for a small subset.
The substantial complexity of the first direction  and  inaccuracy of the second direction  motivate us to derive the explicit expressions of nearest-neighbor and contact  distance distributions of TCP.

In contrast to a homogenous PPP, where the density functions of contact distance and nearest-neighbor distance are identical~\cite{haenggi2012stochastic}, the choice of a reference point  is crucial to characterize  such distance distributions in TCP. In this letter,  we derive the  CDF of distance from a reference point to its nearest point of TCP in three cases: i) the reference point is extraneous to the  TCP, ii) the reference point is \chb{chosen uniformly at random from the TCP}, and iii) a representative cluster is first chosen uniformly at random, and then the reference point is sampled from the representative cluster. We  also derive  closed-form upper \chr{bounds} on the CDFs  in the first two cases.


%
%

\section{System model}
Thomas cluster process is a stationary and isotropic Poisson cluster process generated by a set of offspring points  independently and identically distributed (i.i.d.)  around each point of a parent PPP~\cite{ganti2009interference}. In particular, the locations of parent points are modeled as a homogenous PPP  $\{\bf x\} \equiv \Phi_{\rm p}$ with density $\lambda_{\rm p}$ around which offspring points are distributed according to a symmetric normal distribution with variance $\sigma^2$. Thus, the  probability density function (PDF) of an offspring point location ${\bf y}\in \R^2$ relative to its parent point  is:
\begin{align}\label{eq: PDF normal}
f_{\bf Y}({\bf y})=\frac{1}{2 \pi \sigma^2 }\exp\left(-\frac{\|{\bf y}\|^2}{2 \sigma^2}\right), \quad {\bf y}\in \R^2.
\end{align}
Denote  by ${\cal B}^{\bf x}$ the set of offspring points \chr{for the cluster} centered at ${\bf x}\in \Phi_{\rm p}$. The \chr{cluster} process can be expressed as:
\begin{align}
\Psi=\cup_{{\bf x}\in \Phi_{\rm p}} {\cal B}^{\bf x},
\end{align}
where the number of points per cluster $|{\cal B}^{\bf x}|$ is Poisson distributed random variable with mean $\bar{m}$.  Let ${\cal D }^{{\bf x}} =\{{u: u=\|{\bf x}+{\bf y}\|}, \forall \ {\bf y} \in {\cal B}^{{\bf x}} \}$ be the sequence of distances from the reference point (assumed to be located at the origin) to the points located at the cluster centered at ${\bf x} \in \Phi_{\rm p}$. The elements in the sequence ${\cal D }^{{\bf x}}$ are correlated due to the common factor ${\bf x}$. 
This correlation can be handled by  conditioning on ${\bf x}$  because the locations of offspring points around cluster center are i.i.d. by assumption.  For any choice of the reference point, the conditional distribution of $U$ is Rician with PDF~\cite[Lemma 1]{AfshDhi2015MehrnazD2D1}:
\begin{multline}\label{eq:  Rice PDF}
f_U(u| \|{\bf x}\|)=\mathtt{Ricepdf}(u,\nu=\|{\bf x}\|;\sigma^2)\\
=\frac{u}{ \sigma^2} \exp\left(-\frac{u^2+\nu^2}{2 \sigma^2}\right) I_0\left(\frac{u \nu}{\sigma^2}\right), 
\end{multline}
where  $u$ is a realization of $U$ and $I_0(\cdot)$ is  the modified Bessel function with order zero.

\section{Distance Distributions}
%
%
%

\subsection{Contact distance distribution}
In this subsection, we derive the statistical distribution of the contact distance, which is formally defined next.

\begin{definition}[Contact distance distribution] The contact distance distribution or the empty space distribution function is:
\begin{align}
F_{R_{\rm C}}(r)\triangleq \P(\|\Psi\| \le r)=\P(\Psi({\bf b}( o,r)>0),
\end{align}
\chb{where $R_{\rm C}$ denotes the contact distance}, $\Psi({\bf b}( o,r))$  denotes the number of points within a ball  of radius $r$ centered at  $ o\equiv(0,0)$. Due to the stationarity of TCP, any arbitrary point in $\nbbR^2$ can be treated as reference point or origin. 
\end{definition}
Note that the  reference point, i.e. the origin, is not a part of the  original point process $\Psi$ ($o \notin \Psi$).  The  CDF of  the contact distance  is derived in the next Theorem.
\begin{thm}[Contact distance distribution]\label{Thm: Contact distance} The CDF of the contact distance is:
\begin{multline}\label{eq: exact CDF contact}
 {F}_{R_{\rm C}}(r)=1-\exp\Big(-2 \pi \lambda_{\rm p}\int_0^{\infty} \Big(1- \exp\Big(-\bar{m}\\
 \big(1-Q_1(\frac{v}{\sigma},\frac{r}{\sigma}) \big)\Big) v {\rm d} v \Big)\Big),
\end{multline}
  where $Q_1(\alpha,\beta)$ is the Marcum Q-function defined as $Q_1(\alpha,\beta) = \int_\beta^{\infty} y e^{-\frac{y^2 + \alpha^2}{2}} I_0(\alpha y) {\rm d} y$.
 \end{thm}
\begin{IEEEproof}
 See Appendix~\ref{Proof: Theorem contact}.
\end{IEEEproof}
 
 After characterizing the CDF of  the contact distance, we now \chb{derive a bound on this CDF in the next Corollary.}

 \begin{cor}\label{Cor: bound on contact distribution}
\chb{The CDF $F_{R_{\rm C}}(r)$ can be upper bounded as:}
\begin{align} \label{eq: bound on contact}
 {F}_{R_{\rm C}}(r) \le1-\exp(- \pi \lambda_{\rm p} \bar{m} r^2).
\end{align}
\end{cor}
\begin{IEEEproof}
\chb{Using Taylor expansion of exponential function, we get }
\begin{align*}
 {F}_{R_{\rm C}}(r)& \stackrel{(a)}{\le} 1-\exp\Big(-2 \pi \lambda_{\rm p}\bar{m}\int_0^r  \int_0^{\infty}  f_{U}(u|v) v  {\rm d} v {\rm d} u \Big)\\
&\stackrel{(b)}{=}1-\exp(- \pi \lambda_{\rm p} \bar{m} r^2),
\end{align*}
where $(a)$ follows form  $1-\exp(- \rho x)\le \rho x $, $\rho>0$ and $(b)$ follows from $\int_0^{\infty} f_{U}(u|v) v {\rm d} v=u$.
\end{IEEEproof}
\begin{remark}
\chb{The bound presented in Corollary~\ref{Cor: bound on contact distribution} can be interpreted as the CDF of the contact distance for a homogeneous PPP with the same density $\bar{m}\lambda_P$ as that of the TCP. In fact, it is not uncommon to see this result being used as an ``approximation'' for the true contact distribution of a TCP, e.g., see~\cite{HetPCPGhrayeb2015}. However, to the best of our understanding, this expression has never been claimed to be a bound on the exact CDF. Nevertheless, as will be evident from the numerical results, the bound given by~\eqref{eq: bound on contact}  is rather loose, which highlights the importance of the exact distribution derived in Theorem~\ref{Thm: Contact distance}. }
\end{remark}

\begin{remark} \label{Rem: tightness of contact}
\chb{As per the interpretation discussed in the above remark, the upper bound presented in Corollary 1 is expected to get tighter when TCP itself starts converging towards a homogeneous PPP. This happens when we increase scattering variance $\sigma^2$ or decrease the number of points per cluster $\bar{m}$. Therefore, this bound will be relatively more useful when $\sigma^2$ is large and/or $\bar{m}$ is small.}
\end{remark}

\subsection{Nearest-neighbor distance distribution}
\chb{The contact distance  becomes the nearest-neighbor distance when a reference point is a part of the original point process. }
For any stationary point process, the nearest-neighbor distance is defined as: 
\begin{align}
R_{\rm N}=\|\argmin_{{\bf z}\in\Psi\setminus o }\{\|{\bf z}\|\}\|; \quad  o \in \Psi,
 \end{align}
 where the reference point \chb{is} assumed to be located at the origin. The formal definition of the nearest-neighbor distance distribution is provided next.
 \begin{definition}[Nearest-neighbor distance distribution]  In the stationary point process, the  distribution of nearest-neighbor distance $R_{\rm N}$ is defined as:
\begin{align}
F_{R_{\rm N}}(r)&=1-\P(\Psi({\bf b}( o,r))=1 |  o \in \Psi).
\end{align}
\end{definition}
For derivation of the nearest-neighbor distance distribution, we focus on the following two different cases.

 \begin{itemize}
 
  \item {\em  Case 1}: The reference point is chosen uniformly at random amongst all offspring points. In this case, the  number of offspring points within the reference point's own cluster
 $ N_0^{(1)}$ is {\em number weighted} Poisson distribution, with probability mass function (PMF)~\cite[Sec. 5.3]{chiu2013stochastic}:
\begin{align}\label{eq: number of points PMF V1}
\P(N_0^{(1)}=\ell)=\frac{\ell}{\bar{m}}\frac{\bar{m}^\ell e^{-\bar{m}}}{\ell ! } \quad \text{for } {\ell \in \mathbb{Z}^{+}}\: ,
\end{align} 
 where $\mathbb{Z}^{+}$ is set of positive integer.  \chr{This is because the}  probabilities of different clusters being chosen are proportional to their size (i.e.,  number of offspring points per cluster).  This is a spatial
incarnation of the length-biased sampling  that is the fundamental reason behind the waiting-bus paradox~\cite{feller1968introduction}. In other words, \chb{a point chosen uniformly at random amongst all offspring points is more likely to be from a larger cluster.}

\item {{\em  Case 2}}: 
{\chb{In this case, we choose} the representative cluster  (denoted by ${\cal B}_{0}^{(2)}$) uniformly at random \chr{from the  TCP},
%
and then, given ${\cal B}_{0}^{(2)}$,  the reference point is chosen uniformly at random amongst all offspring points within the representative cluster.} It should be noted that the representative cluster that contains the reference point cannot be empty. Denote by $N_0^{(2)}$ the number of offspring points within the reference point's own cluster. The distribution of $N_0^{(2)}$ is Poisson with mean $\bar{m}$ conditioned on $N_0^{(2)}$  being greater than one, with PMF
\begin{align}\label{eq: number of points PMF V2}
\P(N_0^{(2)}=\ell)=\frac{\bar{m}^\ell e^{-\bar{m}}}{\ell ! (1-e^{-\bar{m}})} \quad \text{for } \ell \in \mathbb{Z}^{+}.
\end{align}
 
  \end{itemize}

%

It is to be noted that with some work, the likelihood function of poisson cluster process presented in~\cite{baudin1981likelihood}  can be used for the derivation of the contact distance and  nearest-neighbor distance  distributions under {\em Case 1}. However, developing sufficient understanding of likelihood functions in order to use them for these derivations require substantial background in measure theory. Therefore, for the wireless network community to comprehend and appreciate these results, we provide an alternate method which circumvents the need to use measure theoretic notions. Moreover,  the CDF of nearest-neighbor distance under {Case 2} cannot be derived using likelihood function of~\cite{baudin1981likelihood}, which made it necessary to develop an alternate approach. 
\chb{Next, we focus on the derivation of the nearest-neighbor distance distribution for the {\em Case 1}.}
\begin{thm} [Case 1: nearest-neighbor distance distribution] \label{Thm: Nearest-neighbor distance V1}The CDF of the nearest-neighbor  distance distribution is: 
\begin{multline}
{F}_{R^{(1)}_{\rm N}}(r)=1-(1-F_{R_{\rm C}}(r)) \int_0^{\infty} \exp\Big(-\bar{m}\ \Big(1-Q_1\big(\frac{v_0}{\sigma},\frac{r}{\sigma}\big)\Big) \Big)\\ f_{V_0}(v_0) {\rm d} v_0,
\end{multline}
where $F_{R_{\rm C}} (\cdot)$ is given by~\eqref{eq: exact CDF contact} and $f_{V_0}(v_0)=\frac{v_0}{ \sigma ^2}\exp\left(-\frac{v_0^2}{2 \sigma^2}\right)$.
\end{thm}
\begin{IEEEproof}
See Appendix~\ref{Proof.  Nearest-neighbor distance V1}. 
\end{IEEEproof}

We further \chb{derive} a closed-form bound on the CDF of the nearest-neighbor distance in the next  Corollary.
\begin{cor}[Case 1: nearest-neighbor distance distribution] \label{Cor: Bound Nearest}
The CDF of nearest-neighbor distance is upper bounded by
\begin{multline}
{F}_{R^{(1)}_{\rm N}}(r)\le 1-\exp(-\pi \lambda_{\rm p} \bar{m} r^2)\\
\times  \exp\Big(-\bar{m}\ \big(1- \exp(-r^2/4\sigma^2) \big)\Big).
\end{multline}
\end{cor}
\begin{IEEEproof}
See Appendix~\ref{Proof: Bound Nearest v1}.
\end{IEEEproof}
As  will be evident from our numerical comparisons, the upper bound presented in Corollary~\ref{Cor: Bound Nearest} tightly approximates the statistics of the nearest-neighbor distance for {\em Case 1}.

We derive the CDF of  the nearest-neighbor distance for {\em Case 2} in the next Theorem.
  \begin{thm} [Case 2: nearest-neighbor distance distribution] \label{Thm: Nearest-neighbor distance v2}The CDF of the nearest-neighbor distribution is: 
\begin{multline}
{F}_{R^{(2)}_{\rm N}}(r)=1-(1-F_{R_{\rm C}}(r)) \int_0^{\infty} \frac{\exp \big(\bar{m} Q_1\big(\frac{v_0}{\sigma},\frac{r}{\sigma}\big)\big)-1}{Q_1\big(\frac{v_0}{\sigma},\frac{r}{\sigma}\big)} \\
 \times \frac{e^{-\bar{m}}}{1-e^{-\bar{m}}} f_{V_0}(v_0) {\rm d} v_0,
\end{multline}
where $F_{R_{\rm C}} (\cdot)$ is given by~\eqref{eq: exact CDF contact} and $f_{V_0}(v_0)=\frac{v_0}{ \sigma ^2}\exp\left(-\frac{v_0^2}{2 \sigma^2}\right)$.
\end{thm}
\begin{IEEEproof}
See Appendix~\ref{Proof: Nearest-neighbor distance v2}. 
\end{IEEEproof}

\begin{figure} 
\center{
{\includegraphics[width=.5\textwidth]{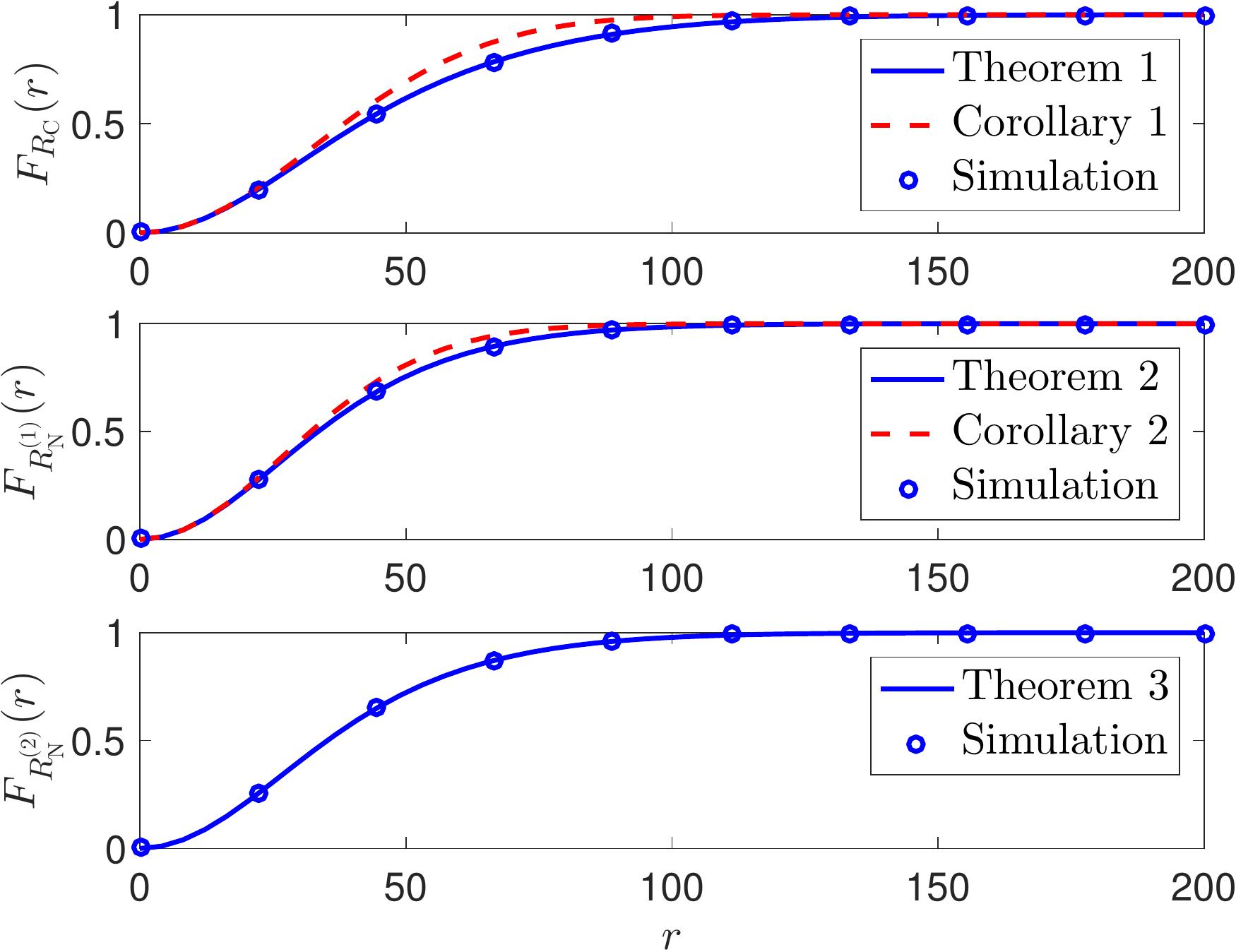}}
}
\caption{CDFs of contact distance and nearest-neighbor distance for the two cases ($\bar{m}=3$, $\sigma=60$, and $\lambda_{\rm p}=50 \times10^{-6}$)}
\label{Fig. Validation}
\end{figure}

We now comment on the  accuracy of the analysis and the tightness of the bounds. As shown in \figref{Fig. Validation},  Theorems \ref{Thm: Contact distance}-\ref{Thm: Nearest-neighbor distance v2} are perfectly  matched with simulation, which corroborates the accuracy of our analysis. In addition, \figref{Fig. Validation} shows that  the upper bound given by Corollary~\ref{Cor: bound on contact distribution} is somewhat loose while Corollary~\ref{Cor: Bound Nearest} provides a  tight upper bound on Theorem~\ref{Thm: Nearest-neighbor distance V1}.
\section{Conclusion}
In this paper, we derived the CDF of contact and nearest-neighbor distance distributions \chb{for a}  TCP.  For the latter, we considered two different approaches of sampling the reference point from the point process.  This work has numerous extensions. From communication perspective, it will enable the characterization of the key performance metrics, e.g., coverage and rate, when the users and/or SBSs are modeled as TCP. From stochastic geometry perspective, it is important to extend the framework to characterize the distribution  
\chr{of the distance of the reference point from its $k^{th}$ closest neighbor.}


%
%

\appendix

\subsection{Proof of Theorem  \ref{Thm: Contact distance}}
\label{Proof: Theorem contact}
 Let us start with the derivation of the probability generating  function (PGF) of the number of point within ${\bf b}(o,r)$. 
 \begin{align}\notag
 &G_{\rm C}(\zz)=\E[\zz^N]=\E\big[\zz^{\sum_{{\bf x}\in \Phi} \sum_{{\bf y}\in {\cal B}^{\bf x}} {\bf 1}\{\|{\bf x}+{\bf y}\|<r\}}\big]\\ \notag
&=\E\Big[\prod_{{\bf x}\in \Phi} \prod_{{\bf y}\in {\cal B}^{\bf x} }\zz^ {{\bf 1}\{\|{\bf x}+{\bf y}\|<r\}}\Big]\\ \notag
&\stackrel{(a)}{=}\E\Big[\prod_{{\bf x}\in \Phi} \exp(-\bar{m} \int_{\R^2}(1-\zz^ {{\bf 1}\{\|{\bf x}+{\bf y}\|<r\}}) f_{\bf Y}({\bf y}){\rm d} {\bf y}\Big]\\ \notag
&\stackrel{(b)}{=}\exp\Big(-\lambda_{\rm p}\int_{\R^2}\Big(1- \exp\Big(-\bar{m} \int_{\R^2}\big(1-\zz^ {{\bf 1}\{\|{\bf z}\|<r\}}\big)\\ \notag
&\times f_{\bf Y}({\bf z}-{\bf x}){\rm d} {\bf z}\Big) {\rm d} {\bf x}\Big)\Big)\\ \notag
&\stackrel{(c)}{=}\exp\Big(-2 \pi \lambda_{\rm p}\int_0^{\infty} \Big(1- \exp\Big(-\bar{m} \int_0^r (1-\zz)\\ \label{eq: PGF contact}
&\times f_{U}(u|v) {\rm d} u\Big) v {\rm d} v \Big) \Big),
 \end{align}
 where $(a)$ follows from the \chr{PGF} of Poisson \chr{random variable} along with the fact that the points in ${\cal B}^{\bf x}$ are i.i.d. with PDF $f_{\bf Y}({\bf y})$ given by~\eqref{eq: PDF normal}, $(b)$ follows from the  probability generating functional (PGFL) of PPP and change of variable ${\bf z}={\bf x}+{\bf y}$, and $(c)$ follows from  converting 
Cartesian to polar coordinates, where $f_{U}(u|v)$ is given by~\eqref{eq:  Rice PDF}. Now, the CDF of contact distance  can be derived as follows:
 \begin{align*}
  & {F}_{R_{\rm C}}(r)=1-\P(N=0)=1-G_{\rm C}(0)
 =1-\\&\exp\Big(-2 \pi \lambda_{\rm p}\int_0^{\infty} \Big(1- \exp\Big(-\bar{m} \int_0^r f_{U}(u|v) {\rm d} u\Big) v {\rm d} v \Big)\Big),
 \end{align*}
 where the final result is obtained by using the definition of Marcum Q-function.

 \subsection{Proof of Theorem~\ref{Thm: Nearest-neighbor distance V1}}
 \label{Proof.  Nearest-neighbor distance V1}
 Denote by ${\cal B}^{(1)}_0$ the set of offspring points within the reference point's own cluster for  {\em Case 1}. The PGF of the number of points within  ${\bf b}( o,r)$ is $G_{\rm N}^{\rm (1)} (\zz)$
\begin{align*}
&{=}\E\Big[\zz^{\sum_{{\bf x}\in \Phi \setminus {\bf x}_0} \sum_{{\bf y}\in {\cal B}^{{\bf x}}} {\bf 1}\{\|{\bf x}+{\bf y}\|<r\} +\sum_{{\bf y} \in {{\cal B}^{(1)}_0\setminus  o}}{\bf 1}\{\|{\bf x}_0+{\bf y}\|<r\} }\Big]\\
&\stackrel{(a)}{=}\E\Big[\prod_{{\bf x}\in \Phi } \prod_{{\bf y}\in {\cal B}^{{\bf x}}} \zz^{{\bf 1}\{\|{\bf x}+{\bf y}\|<r\} } \Big] \E \Big[\prod_{{\bf y} \in {{\cal B}^{(1)}_0\setminus  o}}\zz^{{\bf 1}\{\|{\bf x}_0+{\bf y}\|<r\} }\Big]\\
&\stackrel{(b)}{=}G_{\rm C}(\zz)  \int_0^{\infty}  \sum_{\ell=1}^{\infty}  \frac{\bar{m}^{\ell-1} e^{-\bar{m}}}{(\ell-1)!}  \\
 &\times   \Big(\underbrace{\int_{u=r}^{\infty}  f_{U}(u|v_0) {\rm d} u+\zz \int_{u=0}^{r}  f_{U}(u|v_0) {\rm d} u}_{\rho(\zz,v_0,r)} \Big)^{\ell-1}
f_{V_0}(v_0) {\rm d} v_0\\
&\stackrel{(c)}{=}G_{\rm C}(\zz)  \int_0^{\infty} \exp\Big(- \bar{m} \ \int_0^r (1-\rho(\zz,v_0,r))f_{U}(u|v_0) {\rm d} u\Big) \\
&\times f_{V_0}(v_0) {\rm d} v_0.
\end{align*}
Step (a) follows from the fact that parent point process is a PPP and the offspring point processes are independent of the parent point process, which allows us to handle the reference point's own cluster center separately by Slyvniak's theorem.
Step $(b)$ follows from substituting $G_{\rm C}(\cdot)$ given by~\eqref{eq: PGF contact}, followed by expectation over reference point's own cluster process ${\cal B}^{(1)}_0$, using  PMF given by  \eqref{eq: number of points PMF V1},  followed by converting from Cartesian to polar coordinates by using density function given by~\eqref{eq:  Rice PDF}, where $u=\|{{\bf x}_{0}}+{\bf y}\|$ and $v_0=\|{\bf x}_0\|$. Step $(c)$ follows from PGF of Poisson random variable.  It is to be noted that the position of the reference point relative to its parent point is Gaussian distributed in $\R^2$ and hence random variable $V_0$ with realization $v_0=\|{\bf x}_0\|$ is Rayleigh distributed.
Now, the CDF of  nearest-neighbor distance  can be  obtained by using the fact that ${F}_{R^{(1)}_{\rm N}}(r)= 1-G_{\rm N}^{(1)}(0)$.

 \subsection{Proof of  Corollary \ref{Cor: Bound Nearest}}
 \label{Proof: Bound Nearest v1}
Using the upper bounder derived in Corollary \ref{Cor: bound on contact distribution}, we get
\begin{align*}
{F}_{R^{(1)}_{\rm N}}(r)&\le 1-\exp(- \pi \lambda_{\rm p} \bar{m} r^2) \int_0^{\infty} \exp\Big(-\bar{m}  \\
&\times \int_0^r f_{U}(u|v_0) {\rm d} u\Big) f_{V_0}(v_0) {\rm d} v_0\\
&\stackrel{(a)}{\le} 1-\exp(- \pi \lambda_{\rm p} \bar{m} r^2) \exp\Big(-\bar{m}\\
 &\times \int_0^r \int_0^{\infty}  f_{U}(u|v_0)  f_{V_0}(v_0) {\rm d} v_0 {\rm d} u\Big) 
\end{align*}
where $(a)$ follows from Jensen's inequality. The final result can be obtained by solving $\int_0^r \int_0^{\infty}  f_{U}(u|v_0)  f_{V_0}(v_0) {\rm d} v_0=1-\exp\left(-\frac{r^2}{4 \sigma^2}\right)$.

\subsection{Proof of Theorem~\ref{Thm: Nearest-neighbor distance v2}}
\label{Proof: Nearest-neighbor distance v2}
Denote by ${\cal B}^{(2)}_0$ the set of offspring points within the reference point's own cluster for  {\em Case 2}. The PGF of the number of points within  ${\bf b}( o,r)$ is $G_{\rm N}^{(2)} (\zz)$

\begin{align*}
&=\E\Big[\zz^{\sum_{{\bf x}\in \Phi \setminus {\bf x}_0} \sum_{{\bf y}\in {\cal B}^{{\bf x}}} {\bf 1}\{\|{\bf x}+{\bf y}\|<r\} +\sum_{{{\bf y} \in {\cal B}^{(2)}_0\setminus  o}}{\bf 1}\{\|{\bf x}_0+{\bf y}\|<r\} }\Big]\\
&\stackrel{(a)}{=}G_{\rm C}(\zz)  \int_0^{\infty}  \sum_{\ell=1}^{\infty} \Big(\int_{u=r}^{\infty}  f_{U}(u|v_0) {\rm d} u+\zz \int_{u=0}^{r}  f_{U}(u|v_0) {\rm d} u \Big)^{\ell-1}\\
 &\times  \frac{\bar{m}^\ell e^{-\bar{m}}}{\ell! (1-e^{-\bar{m}})}
f_{V_0}(v_0) {\rm d} v_0\\
&{=}G_{\rm C}(\zz)  \int_0^{\infty} \frac{\exp(\bar{m}\rho(\zz,v_0,r))-1}{\rho(\zz,v_0,r)}  \frac{e^{-\bar{m}}}{1-e^{-\bar{m}}} f_{V_0}(v_0) {\rm d} v_0
\end{align*}
where $(a)$ follows on the same lines as the proof of Theorem~\ref{Thm: Nearest-neighbor distance V1}, using the PMF of  number of points within reference point's own cluster given by~\eqref{eq: number of points PMF V2}.  Now, the   CDF of  nearest-neighbor distance  for {\em Case 2} is
 ${F}_{R^{(2)}_{\rm N}}(r)= 1-G_{\rm N}^{(2)}(0)$.

\bibliographystyle{IEEEtran}
\bibliography{Letter_Thomas_v8_arXiv.bbl}

\begin{thebibliography}{10}
\providecommand{\url}[1]{#1}
\csname url@samestyle\endcsname
\providecommand{\newblock}{\relax}
\providecommand{\bibinfo}[2]{#2}
\providecommand{\BIBentrySTDinterwordspacing}{\spaceskip=0pt\relax}
\providecommand{\BIBentryALTinterwordstretchfactor}{4}
\providecommand{\BIBentryALTinterwordspacing}{\spaceskip=\fontdimen2\font plus
\BIBentryALTinterwordstretchfactor\fontdimen3\font minus
  \fontdimen4\font\relax}
\providecommand{\BIBforeignlanguage}[2]{{%
\expandafter\ifx\csname l@#1\endcsname\relax
\typeout{** WARNING: IEEEtran.bst: No hyphenation pattern has been}%
\typeout{** loaded for the language `#1'. Using the pattern for}%
\typeout{** the default language instead.}%
\else
\language=\csname l@#1\endcsname
\fi
#2}}
\providecommand{\BIBdecl}{\relax}
\BIBdecl

\bibitem{chiu2013stochastic}
S.~N. Chiu, D.~Stoyan, W.~S. Kendall, and J.~Mecke, \emph{Stochastic Geometry
  and its Applications}, 3rd~ed.\hskip 1em plus 0.5em minus 0.4em\relax New
  York: John Wiley and Sons, 2013.

\bibitem{SahaAfshDh2016}
C.~Saha, M.~Afshang, and H.~S. Dhillon, ``Enriched {$K$}-tier {HetNet} model to
  enable the analysis of user-centric small cell deployments,''
  \emph{\emph{submitted to} IEEE Trans. on Wireless Commun.}, 2016, available
  online:arxiv.org/abs/1606.06223.

\bibitem{3gpp2}
3GPP, ``{Consideration of {UE} Cluster Position and {PeNB} {TX} Power in
  Heterogeneous Deployment Configuration 4},'' Discussion/ Decision
  {R1-100477}, Jan. 2010, 8.2.1 Relevant scenarios of Heterogeneous Networks.

\bibitem{spatialmodelingAndrews2010}
J.~G. Andrews, R.~K. Ganti, M.~Haenggi, N.~Jindal, and S.~Weber, ``A primer on
  spatial modeling and analysis in wireless networks,'' \emph{IEEE Commun.
  Magazine}, vol.~48, no.~11, pp. 156--163, Nov. 2010.

\bibitem{haenggi2012stochastic}
M.~Haenggi, \emph{Stochastic Geometry for Wireless Networks}.\hskip 1em plus
  0.5em minus 0.4em\relax Cambridge University Press, 2012.

\bibitem{baudin1981likelihood}
M.~Baudin, ``Likelihood and nearest-neighbor distance properties of
  multidimensional poisson cluster processes,'' \emph{Journal of Applied
  Probability}, pp. 879--888, 1981.

\bibitem{HetPCPGhrayeb2015}
Y.~J. Chun, M.~O. Hasna, and A.~Ghrayeb, ``Modeling heterogeneous cellular
  networks interference using poisson cluster processes,'' \emph{IEEE Journal
  on Sel. Areas in Commun.}, vol.~33, no.~10, pp. 2182--2195, Oct. 2015.

\bibitem{Wang2016}
Y.~Wang and Q.~Zhu, ``Modeling and analysis of small cells based on clustered
  stochastic geometry,'' \emph{IEEE Commun. Letters}, 2016, to appear.

\bibitem{ganti2009interference}
R.~K. Ganti and M.~Haenggi, ``Interference and outage in clustered wireless ad
  hoc networks,'' \emph{IEEE Trans. on Info. Theory}, vol.~55, no.~9, pp.
  4067--4086, Sep. 2009.

\bibitem{AfshDhi2015MehrnazD2D1}
M.~Afshang, H.~S. Dhillon, and P.~H.~J. Chong, ``Modeling and performance
  analysis of clustered device-to-device networks,'' \emph{IEEE Trans. on
  Wireless Commun.}, vol.~15, no.~7, pp. 4957--4972, Jul. 2016.

\bibitem{feller1968introduction}
W.~Feller, \emph{An introduction to probability theory and its
  applications}.\hskip 1em plus 0.5em minus 0.4em\relax John Wiley \& Sons
  London-New York-Sydney-Toronto, 1968, vol.~3.

\end{thebibliography}

\end{document}